\documentclass[preprint,noshowpacs,amsmath,amssymb,superscriptaddress]{revtex4}
\usepackage[dvips]{graphicx}
\usepackage{amsmath}
\usepackage{bm}

\begin{document}

\newcommand{\be}{\begin{equation}}
\newcommand{\ee}{\end{equation}}

\title{Neutrino masses and mixing: Singular mass matrices and Quark-lepton
symmetry}

\author{I.~Dorsner}
\email{idorsner@ictp.trieste.it}
\affiliation{International Centre
for Theoretical Physics, Strada Costiera 11, 31014 Trieste, Italy}
\author{A.~Yu.~Smirnov}
\email{smirnov@ictp.trieste.it}
\affiliation{International Centre
for Theoretical Physics, Strada Costiera 11, 31014 Trieste, Italy}
\affiliation{Institute for Nuclear Research, Russian Academy of
Sciences, Moscow, Russia}


\begin{abstract}
We suggest an approach to explain the observed pattern of the
neutrino masses and mixing which employs the weakly violated
quark-lepton equality and does not require introduction of an {\it
ad hoc\/} symmetry of the neutrino sector. The mass matrices are
nearly equal for all quarks and leptons. They have very small
determinant and hierarchical form with expansion parameter
$\lambda \sim \sin \theta_c \sim \sqrt{m_{\mu}/m_{\tau}}$. The
latter can be realized, e.g., in the model with $U(1)$ family
symmetry. The equality is violated at the $\sim \lambda^2$ level.
Large lepton mixing appears as a result of summation of the
neutrino and charged lepton rotations which diagonalize
corresponding mass matrices in contrast with the quark sector
where the up quark and down quark rotations cancel each other. We
show that the flip of the sign of rotation in the neutrino sector
is a result of the seesaw mechanism which also enhances the
neutrino mixing. In this approach one expects, in general,
deviation of the 2-3 mixing from maximal, $s_{13} \sim
(1~$--$~3)\lambda^2$, hierarchical neutrino mass spectrum, and
$m_{ee} < 10^{-2}$\,eV. The scenario is consistent with the
thermal leptogenesis and (in SUSY context) bounds on lepton number
violating processes, like $\mu \rightarrow e \gamma$.
\end{abstract}

\maketitle

\section{Introduction}

One of the main results in the neutrino physics is a surprising
pattern of the lepton mixing which differs substantially from the
quark mixing pattern. The 2-3 leptonic mixing is maximal or nearly
maximal, the 1-2 mixing is large but not maximal and the 1-3
mixing is small or very small (see
\cite{Giunti:2003cq,Smirnov:2003xe} for recent reviews). No
apparent regularities or relations between mixing parameters as
well between mass ratios of different fermions have been found,
except for probably accidental relation $\theta_{12} + \theta_C =
\theta_{23} \sim 45^\circ$. Furthermore, the data on masses and
mixings show some degree of ``chaoticity''.

In this connection, there are two essential issues on the way to
the underlying physics:

\begin{itemize}

\item

Quark-lepton symmetry: Is it still realized at some level? and

\item

New symmetry of Nature behind neutrino masses and mixings: Does it
exist?

\end{itemize}

As is well known, the exact  quark-lepton symmetry is violated by
difference of masses of quarks and charged leptons of the first
and second generations. It seems that neutrino mixing further
deepens this difference.

On the other hand, several features of the neutrino data indicate
certain symmetry which is not realized in the quark and charged
lepton sectors (we will call it the ``neutrino symmetry"):

- Maximal (or near maximal) 2-3 mixing;

- Small 1-3 mixing: the fact that

\be \sin \theta_{13} \ll \sin \theta_{12} \times \sin \theta_{23}
\ee

indicates some special structure of the mass matrix;

- Possible quasi-degenerate neutrino mass spectrum. This  is
hinted by (i) a general consideration in physics according to
which large mixing is associated with  degeneracy, ii) the
neutrinoless beta decay
result~\cite{HM-neg1,HM-pos,vissani,HM-neg2,Klapdor-Kleingrothaus:2004ge},
(iii) the cosmological analysis which uses particular set of
observations~\cite{cosm3} (see however~\cite{cosm,cosm1,cosm2}).

These features can be related. The same symmetry can lead to the
maximal 2-3 mixing and zero 1-3 mixing. So, breaking of the
symmetry will generate simultaneously the non-zero 1-3 mixing and
deviation of the 2-3 mixing from maximal value. (See however
\cite{Low:2004wx}.) Maximal mixing can be related to
the quasi-degenerate mass spectrum, {\it etc}.\\

There is a number of studies which explore various ``neutrino
symmetries'' like $Z_2$, $A_4$~\cite{A4} or
$SO(3)$~\cite{so3,Antusch:2004xd,Antusch:2004re} (see
\cite{Altarelli:2003vk} for review). Apparently these symmetries
being exact or approximate cannot be extended to the charged
lepton sector where the hierarchy of masses, and in particular,
inequality $m_{\mu} \ll m_{\tau}$, exists. Even more difficult is
to include in the same scheme quarks which show small mixings.
Realization of ``neutrino symmetries'' usually requires
introduction of (i) new leptons and quarks, (ii) complicated Higgs
sector to break the symmetry, (iii) additional symmetries to
forbid unwanted couplings associated to new fermions and scalars,
{\it etc}. Thus, in the ``neutrino symmetry'' scenario the
observed pattern of mixing has profound
implications and requires substantial extensions of known structures. \\

It is not excluded, however, that the ``neutrino  symmetry'' is
just misleading interpretation. In fact, till now the only solid
indication of the new symmetry is the maximal 2-3 mixing. Notice,
however that $\sin^2 2\theta_{23} = 1$ is obtained as the best fit
point in the $2\nu$ analysis of the atmospheric neutrino data
~\cite{atm}. At 90 \% C.L. $\sin^2 2\theta_{23} > 0.9$~\cite{atm}.
The K2K experiment gives even weaker bound on 2-3
mixing~\cite{K2K}. Furthermore, $\sin^2 2\theta_{23}$ is a bad
quantity to describe the deviation of mixing from maximal. From
theoretical point of view the relevant parameter would be
\begin{equation}
D_{23} \equiv 1/2 - \sin^2 \theta_{23}. \label{devia}
\end{equation}
Then the present experimental bound on the deviation is \be
|D_{23}| < 0.15~~~ (90\% {\rm C.L.}). \ee That is, $|D_{23}|
\sim\sin^2 \theta_{23}$ is still possible and at the moment we
cannot say that the 2-3 mixing is really near maximal one.
Moreover, the latest analysis, of the atmospheric neutrino data
(without renormalization of the original fluxes) shows some excess
of the $e$-like events at low energies (the sub-GeV events) and
the absence of excess in the multi-GeV sample. This gives a hint
of non-zero $D_{23}$~\cite{atm}. The deviation can show up in the
generic $3\nu$ analysis of the data with the solar oscillation
parameters
taken into account.\\

In this connection we will explore an opposite ``no-neutrino
symmetry'' approach which does not rely on a special symmetry for
the neutrino sector. In contrast, we will employ the quark-lepton
symmetry as much as possible.

Some elements of our approach have already been considered before.

We use the mass matrix structure which leads to mixing angles of
the order \be \tan \theta_{ij} \sim \sqrt{\frac{m_i}{m_j}},
\label{lambda} \ee where $m_i$ are the
eigenvalues~\cite{Weinberg:hb,Wilczek:uh,Fritzsch:1977za}.

The enhancement of lepton mixing is a result of summation of
rotations which diagonalize the neutrino and charged lepton mass
matrices~\cite{barshay}. In contrast, the rotations cancel each
other in the quark sector thus leading to small quark mixing. In
this case the atmospheric mixing angle equals \be \theta_{23} \sim
\sqrt{\frac{m_2}{m_3}} + \sqrt{\frac{m_{\mu}}{m_{\tau}}}. \label{}
\ee The ratio of neutrino masses is bounded from below by mass
squared differences measured in the solar $(\Delta m^2_{12})$ and
the atmospheric $(\Delta m^2_{23})$ neutrino experiments: \be
\frac{m_2}{m_3} \geq \sqrt{\frac{\Delta m^2_{12}}{\Delta
m^2_{23}}} = 0.18^{+ 0.22}_{-0.08}. \label{eq:hie1} \ee The
corresponding mass ratio for the charged leptons is smaller:
$m_\mu/m_{\tau} \approx 0.06$. Even taking equality in
(\ref{eq:hie1}) (which would correspond to the hierarchical mass
spectrum) we find $\theta_{23} \sim 38^0$ which  is well within
the allowed region.

We employ the seesaw mechanism~\cite{sees} and partial seesaw
enhancement of the neutrino mixing \cite{senhan}.

We also posit a symmetric form for the mass matrix structure. It
is in the case of symmetric matrices that the strong mass
hierarchy and large mixing can be reconciled provided that the
determinant of matrix is very small.

Finally, in the ``democratic approach'' the idea that to leading
approximation all the mass matrices in the lepton sector are
proportional to each other has been pursued in
\cite{Akhmedov:2000yt}. It has been further
extended to the quark sector as well in \cite{Dermisek:2003rw}.\\

The paper is organized as follows. In Section~\ref{approach} we
formulate our ``no-neutrino symmetry'' approach. In
Section~\ref{singular} we describe main features of the mass
matrices and find masses and mixing angles. In
Section~\ref{predictions} we obtain generic predictions of the
approach. In Section~\ref{discussion} we consider the theoretical
implications. Conclusions follow in Section~\ref{conclusions}.
Numerical results are presented in the Appendix.

\section{No-neutrino symmetry approach}
\label{approach}

In what follows we assume the following.

\noindent 1). The weakly broken quark-lepton symmetry  is realized
in terms of the mass matrices and not in terms of observables
(masses and mixing angles). The Yukawa couplings for all quarks
and leptons are nearly equal, so that the matrices of the
couplings can be written as \be \hat{Y}_K \approx \hat{Y}_0 +
\delta \hat{Y}_K, ~~~ K = u, d, l, D. \label{yukmat} \ee Here
index $D$ refers to the Dirac type matrix of neutrinos. The
dominant structure is given by $\hat{Y}_0$ which is common for all
fermions, whereas the matrices of small corrections, $\delta
\hat{Y}_K$, are different for different fermions. The smallness of
$\delta \hat{Y}_K$ can be specified in two different ways which
have different theoretical implications: \be (\delta
\hat{Y}_K)_{ij} \ll (\hat{Y}_0)_{ij}, \label{first} \ee that is,
the relative corrections are small to all matrix elements, or \be
(\delta \hat{Y}_K)_{ij} \ll 1, \label{second} \ee if the largest
element, $(\hat{Y}_0)_{33}$, is normalized to 1. In what follows
for definiteness we will elaborate on the first possibility.\\

\noindent 2). We assume that the matrix $\hat{Y}_0$ is singular:
whole matrix $\hat{Y}_0$ as well as the sub-matrices 2-3 and 1-3
have zero (very small) determinants. As a consequence, $\hat{Y}_0$
is ``unstable" in a sense that small  perturbations, $\delta
\hat{Y}_K$, lead to significant difference in the eigenvalues
(masses) and eigenstates (mixings). This allows us to explain (see
Section~\ref{singular}) substantial deviation from the
quark-lepton symmetry at the level of observables.

In what follows we consider the following symmetric singular
structure
\begin{equation}
\hat{Y}_0 =   \left(
\begin{tabular}{ccc}
$\lambda^4~$ & $~\lambda^3~$ & $~\lambda^2$\\
$\lambda^3$ & $\lambda^2$ & $\lambda$\\
$\lambda^2$ & $\lambda$ & 1\\
\end{tabular}
\right),
\label{yzero}
\end{equation}
where the expansion parameter \be \lambda \sim \sin\theta_c \sim
\mbox{0.2~--~0.3}. \ee We will comment on other possibilities in
Section~\ref{discussion}.\\

\noindent 3). The smallness of neutrino mass is explained by the
seesaw mechanism~\cite{sees}:
\begin{equation}
\hat{m}_{\nu} = - \hat{m}_D \hat{M}_R^{-1} \hat{m}_D^T~~,
\label{eq:seesaw}
\end{equation}
where $\hat{m}_D \equiv \hat{Y}_{D} v_{1}$ is the Dirac mass
matrix, and $v_{1}$ is the electroweak vacuum expectation value
(VEV) which generates masses of the upper fermions. The seesaw
type II contribution, if exists, is small and can contribute to
the correction matrix $\delta \hat{Y}_K$.

For simplicity we assume that mass matrix of the right-handed (RH)
neutrinos, $\hat{M}_R$, has the same structure as given in
Eqs.~(\ref{yukmat}) and (\ref{yzero}). This could correspond to a
situation when all fermionic components are in the same multiplet
and the flavor information is in fermions, whereas Higgs
multiplets are flavorless. In general, this is not necessary,
since the RH neutrino mass matrix has different gauge properties
and is generated  by  different Higgs multiplet VEV.
Also it may have different expansion parameter. \\

The seesaw mechanism plays the triple role here. (i) It explains
smallness of the neutrino mass. (ii) It flips the sign of rotation
which diagonalizes the light neutrino mass matrix, so that in the
lepton sector the up and down rotations sum up (in contrast to the
quark sector) thus leading to large lepton mixing. (iii) It
enhances moderately (by factor of $\sim 2$) the mixing angles
which come from the neutrino mass matrix. The last two
facts---flipping of the relative sign of rotations and the
moderate seesaw enhancement of the neutrino mixing
angle---lead to large lepton mixings.\\

The situation is different in the quark sector. The same
dominant form for the mass matrices of the up and down quarks
leads due to cancellation of rotations to zero mixing equal to the
identity matrix. The CKM matrix originates from the mismatch between
correction matrices $\delta \hat{Y}_u$ and $\delta \hat{Y}_d$ which
appear small in our approach. This in turn guarantees the
smallness of the CKM angles.\\

We parametrize the complete matrix of Yukawa coupling
(\ref{yukmat}) as
\begin{equation}
\label{ytotal} \hat{Y}_K =  \left(
\begin{tabular}{ccc}
$(1 + \epsilon_{11}^K)\lambda^4~$ & $~(1 +
\epsilon_{12}^K)\lambda^3~$ &
$~(1 + \epsilon_{13}^K)\lambda^2$\\
$(1 + \epsilon_{12}^K)\lambda^3$ & $(1 + \epsilon_{22}^K)\lambda^2$ &
$(1 + \epsilon_{23}^K)\lambda$\\
$(1 + \epsilon_{13}^K)\lambda^2$ & $(1 + \epsilon_{23}^K)\lambda$ & 1\\
\end{tabular}
\right) y_K,
\end{equation}
where the range for the corrections, $\epsilon_{ij}$, is
restricted by $\lambda$: \be |\epsilon_{ij}^K| \leq \lambda, ~~~K
= u,d, l, D, M, \label{eps} \ee for all $i,j$ in the first case
(\ref{first}). The overall multipliers, $y_K \simeq 1$, describe
the amount of non-unification of the third generation of quarks
and leptons. They can also be introduced as the corrections to 33
elements: $1\rightarrow(1+\epsilon_{33}^K)$.

The mass matrices (without renormalization group effects) equal:
\be
\begin{array}{ll}
\hat{m}_{K} =  \hat{Y}_K v_1, & ~~K = u, D,\\
\hat{m}_{K} =  \hat{Y}_K v_2, & ~~K = d, l,\\
\hat{M}_{R} =  \hat{Y}_K M_0, & ~~K = M.
\end{array}
\ee Here $v_1$ and $v_2$ are the VEVs of the two Higgs doublets
and $M_0$ is the overall scale of RH neutrino masses.

In what follows we will consider for simplicity $\epsilon_{ij}$ to
be real.

\section{Singular mass matrices, masses and mixings}
\label{singular}

\subsection{Expansion parameter}

The value of expansion parameter is determined essentially by the
condition (\ref{eps}) which encodes degree of violation of the
quark-lepton symmetry in our approach and by the ratio of muon to
tau lepton masses which shows the weakest mass hierarchy.
According to (\ref{ytotal}) we obtain \be \label{mutotau}
\frac{m_\mu}{m_\tau} \approx \lambda^2 (\epsilon_{22}^l -
2\epsilon_{23}^{l}) \sim \lambda^2 \cdot \epsilon \leq \lambda^3,
\ee where, in general, by $\epsilon$ we will denote combinations
of $\epsilon_{ij}$ of the order $\epsilon_{ij}$.

There are two solutions of Eq.~(\ref{mutotau}) depending on the
sign of the mass ratio. If $m_\mu/m_\tau < 0$, the smallest value
of $\lambda$ would correspond to $\epsilon_{22}^l \sim - \lambda$
and $\epsilon_{23}^l = \lambda$, so that the ratio equals
$3\lambda^3$, and consequently, \be \lambda \sim \left(-
\frac{m_\mu}{3 m_\tau}\right)^{1/3}. \ee Using value of the mass
ratio at the GUT scale, $m_{\mu}/m_{\tau} = 0.045$, we obtain
$\lambda \geq 0.26$. In this case the corrections enhance the
mixing: \be \tan 2\theta_{23}^l = \frac{2(1 +  \epsilon_{23}^l)
\lambda}{1 - (1+\epsilon_{22}^l) \lambda^2 } \approx \frac{2 (1 +
\lambda) \lambda}{1 - \lambda^2 }. \ee For $m_\mu/m_\tau > 0$ the
smallest $\lambda$ corresponds to $\epsilon_{22}^l \sim \lambda$
and $\epsilon _{23}^l = - \lambda$. The required value of
$\lambda$ is approximately the same but the mixing is smaller.

Notice that $\lambda = \sin \theta_c = 0.22$ would require
$\epsilon_{22}^l - 2\epsilon_{23}^{l} = 1~$--$~2$, that is, large
corrections.

\begin{figure}
\begin{center}
\includegraphics[width=4in]{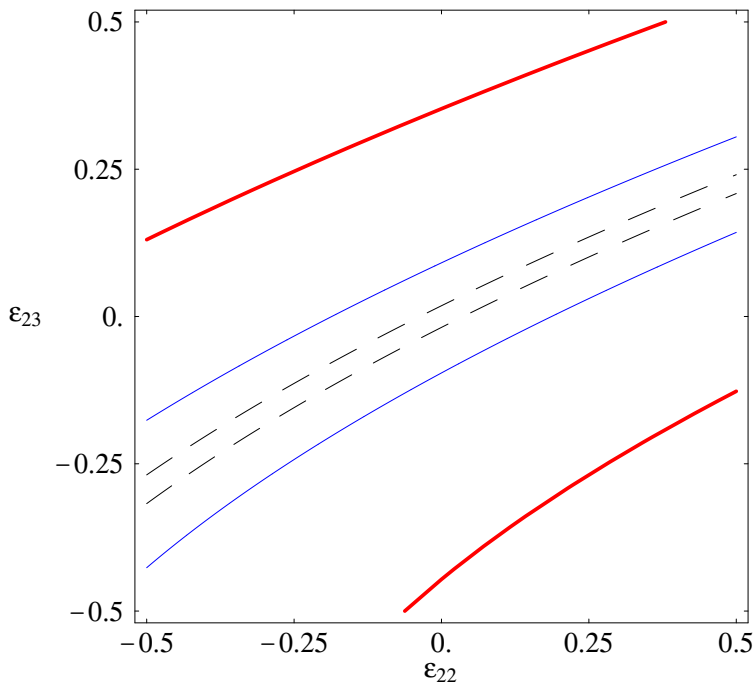}
\end{center}
\caption{\label{contour} The lines of constant ratio
$m_\mu/m_\tau=0.045$ (thick solid line), $m_s/m_b=0.011$ (thin
solid line), and $m_c/m_t=0.0022$ (dashed line) in
$\epsilon_{22}$-$\epsilon_{23}$ plane at the GUT scale.}
\end{figure}

Stronger mass hierarchy of quarks can be obtained taking values of
$\epsilon_{22}$ and $\epsilon_{23}$ closer to zero. In
Fig.~\ref{contour} we show the lines of constant mass ratios
$m_2/m_3$ in the $\epsilon_{22}$-$\epsilon_{23}$ plane for the
quarks and charged leptons. The figure indicates certain hierarchy
of the ${22}$ and ${23}$ corrections: $\epsilon^{u} \ll
\epsilon^{d} \ll \epsilon^{l}$. However, this hierarchy cannot be
established for all matrix elements due to the need to reproduce
observed mixing angles. In particular, value of the 2-3 CKM mixing
still prevents the deviations in the 2-3 sector of the up and down
quarks from being extremely small simultaneously.

Explanation of other observables, especially in the 1-2 sectors,
requires that some $\epsilon^{u},~ \epsilon^{d} \sim \lambda$ (see
the Table~\ref{tab:table2} in the Appendix).

\subsection{Masses and mixing from $\hat{Y}_K$}
\label{YK}

The matrix $\hat{Y}_0$ can be diagonalized by $U_{0} = U_{23}
U_{13}$, where the corresponding rotation angles equal $\tan
\theta_{23}  = \lambda + O(\lambda^3)$ and $\tan \theta_{13}  =
\lambda^2 + O(\lambda^4)$. After these rotations the 1-2 matrix
becomes zero and therefore masses and mixing of the first and
second generations are determined completely by the corrections to
$\hat{Y}_0$. In fact, only 33 elements of matrices are nonzero and
we will call this basis the ``33'' basis.

Formally one could work immediately in the ``33'' basis. In this
basis however there is no guideline (apart from the experimental
data) how corrections should be introduced. One can consider the
matrix (\ref{ytotal}) as an ansatz for introduction of the
corrections. It by itself leads to certain qualitative pattern of
masses and mixing though quantitative predictions depend
substantially on particular values of $|\epsilon|$'s within
interval $(0~$--$~\lambda)$. Furthermore the ansatz (\ref{ytotal})
has certain theoretical implications which we will outline in
Section~\ref{discussion}. For a different ansatz where the
dominant structure of the Yukawa matrices has a democratic form
and related phenomenological considerations see
\cite{Akhmedov:2000yt,Dermisek:2003rw}.

In what follows we find the parametric expressions for the
observables in terms of $\lambda$ and $\epsilon_{ij}^K$. We
discuss then restrictions on $\epsilon_{ij}^K$ and relations
between them. The detailed study of $\epsilon_{ij}^K$ and their
possible origins
will be given elsewhere~\cite{DS}.\\

The complete mass matrix $\hat{Y}_K$ can be diagonalized with high
accuracy by three successive rotations: $U = U_{23} U_{13}
U_{12}$. The 2-3 rotation is determined by \be \sin \theta_{23}
\approx \lambda (1 + \epsilon_{23}), \label{angle23} \ee and the
1-3 rotation---by \be \sin \theta_{13} \approx \lambda^2 (1 +
\epsilon_{13}). \label{angle13} \ee Here we omit the superscript
for $\theta_{ij}$ and $\epsilon_{ij}$, since these results apply
to all fermions.

As a result of these two rotations we find the mass of the
heaviest eigenstate \be m_3 = 1 + \lambda^2 + O(\lambda^2
\epsilon), \ee and the matrix for the first and second
generations:
\begin{equation}
\hat{m}_{12} =
\lambda^2 \left(
\begin{tabular}{cc}
$\lambda^2 (\epsilon_{11} - 2\epsilon_{13})~$ & $~\lambda
(\epsilon_{12}
- \epsilon_{13} - \epsilon_{23})$\\
$\lambda (\epsilon_{12}- \epsilon_{13} - \epsilon_{23})~$ &
 $\epsilon_{22} - 2\epsilon_{23}$ \\
\end{tabular}
\right), \label{mat12}
\end{equation}
where each matrix element is given in the lowest order in
$\epsilon$. Diagonalization of (\ref{mat12}) gives \be \tan
2\theta_{12} = 2 \lambda \frac{ (\epsilon_{12}- \epsilon_{13} -
\epsilon_{23})} {\epsilon_{22} - 2\epsilon_{23} + O(\lambda^2
\epsilon)}=\lambda \cdot r(\epsilon), \label{angle12} \ee and
masses of the lightest fermions
\begin{subequations}
\begin{eqnarray}
m_2 & = &  \lambda^2 (\epsilon_{22} - 2\epsilon_{23}) = \lambda^2
\cdot \epsilon \leq \lambda^3,
\\\nonumber
\\
\label{mass1}
m_1 & = & \lambda^4 \left[ \epsilon_{11} -
2\epsilon_{13} - \frac{(\epsilon_{12} - \epsilon_{13} -
\epsilon_{23})^2}{\epsilon_{22} - 2 \epsilon_{23}}
\right]=\lambda^4 \cdot \epsilon.
\end{eqnarray}
\end{subequations}
Here \be r(\epsilon) \equiv \frac{\epsilon_1}{ \epsilon_2}\ee
where $\epsilon_1$, $\epsilon_2$ are functions of the order
$\epsilon$ and parametrically $r(\epsilon) = O(1)$. However,
strong cancellation can occur in $\epsilon_i$. Also in some cases
different terms in $\epsilon_i$ can sum up producing an
enhancement. As a result, the ratio can be in rather wide range
$r(\epsilon) \sim 10^{-1}~$--$~10$.

Notice that the lightest mass is of the order $\lambda^4 \cdot
O(\epsilon) \leq \lambda^5 \sim 10^{-3}$ which gives correct order
of magnitude for the down quarks and charged leptons.

The scenario predicts the following hierarchy of masses: \be
\frac{m_2}{m_3} = \lambda^2 \epsilon, ~~~ \frac{m_1}{m_2} =
\lambda^2 r(\epsilon), ~~~ \frac{m_1}{m_3} = \lambda^4 \epsilon.
\label{massratios} \ee

The experimental values of mass ratios, $m_2^K/m_3^K$ and
$m_1^K/m_3^K$, can be obtained provided that the combinations of
$\epsilon$ in (\ref{massratios}) take on the values given in the
Table~\ref{tab:table4}.
\begin{table}[h]
\caption{\label{tab:table4} The values of combinations of
$\epsilon$ in (\ref{massratios}) that yield correct values of the
mass ratios at the GUT scale. We take $\lambda=0.26.$}
\begin{ruledtabular}
\begin{tabular}{clll}
 & $\epsilon_u$ & $\epsilon_d$ & $\epsilon_l$ \\
  \hline
  $m_2/m_3$ & 0.032 & 0.16 & 0.66 \\
  $m_1/m_3$ & 0.0010 & 0.14 & 0.047 \\
\end{tabular}
\end{ruledtabular}
\end{table}
So, cancellation or enhancement in the combinations of $\epsilon$
is needed which testifies that certain relations or/and hierarchy
between $\epsilon_{ij}^K$ exist. Random selection of parameters
$|\epsilon_{ij}^K|$ in the intervals $(0~$--$~\lambda)$ will not
produce correct values of masses in most of the cases. The
observables are very sensitive to choice of $\epsilon$. It is this
high sensitivity to $\epsilon$ that produces substantially
different masses of up and down quarks and leptons.

Notice that according to (\ref{mat12}) and (\ref{mass1}) both
$m_1$ and 1-2 mixing will be enhanced if
$\epsilon_{22} \approx 2\epsilon_{23}$.\\

The physical mixing matrix is a mismatch of the left rotations
which diagonalize the mass matrices of the up and the down
components of the weak doublets: $U = U(up)^{\dagger} U(down).$
Since the mass matrices of the up and down fermions are very
similar, especially in 2-3 sector, they are diagonalized by rather
similar rotations. In particular, the angles of up and down
rotation have the same sign thus cancelling each other in the
physical mixing matrix, so that  $U \sim I$. This explains the
smallness of the quark mixing angles. In contrast, due to the
seesaw the neutrino rotation may flip the sign, so that the
rotations in lepton sector will sum up leading to large mixing
angles.

\subsection{Quark mixing}

The CKM matrix is given by \be V_{CKM} =
U_{12}^{u~\dagger}U_{13}^{u~\dagger}U_{23}^{u~\dagger} U_{23}^d
U_{13}^d U_{12}^d.\ee Using Eqs.~(\ref{angle23}), (\ref{angle13}),
and (\ref{angle12}) we obtain the elements of the CKM matrix in
the leading order in $\lambda$ and $\epsilon$:
\begin{subequations}
\begin{eqnarray}
V_{cb} &\cong& \lambda (\epsilon^d_{23}-\epsilon^u_{23}) = \lambda
\cdot \epsilon,
\\\nonumber
\\
V_{ub} &\cong& \lambda^2 \left(\epsilon^d_{13}-\epsilon^u_{13} -
\frac{(\epsilon^d_{23}-\epsilon^u_{23})(\epsilon^u_{12}-\epsilon^u_{13}-
\epsilon^u_{23})}{(\epsilon^u_{22}-2 \epsilon^u_{23})} \right) =
\lambda^2 \cdot \epsilon,
\\\nonumber
\\
V_{us} &\cong& \lambda \left(
\frac{(\epsilon^d_{12}-\epsilon^d_{13}-
\epsilon^d_{23})}{(\epsilon^d_{22}-2 \epsilon^d_{23})} -
\frac{(\epsilon^u_{12}-\epsilon^u_{13}-
\epsilon^u_{23})}{(\epsilon^u_{22}-2 \epsilon^u_{23})}\right) =
\lambda \cdot r(\epsilon).
\end{eqnarray}
\end{subequations}
These elements have correct order of magnitude without any need
for some special correlation between $\epsilon^K_{ij}$. Indeed,
for $\lambda=0.26$, $V_{cb}$ requires $\epsilon=0.12 \approx 0.46
\lambda$, $V_{ub}$: $\epsilon = 0.042 \approx 0.16 \lambda$, and
$V_{us}$: $r(\epsilon)\approx 0.86$.

The hierarchy of the quark mixings is naturally reproduced: \be
V_{us} \sim \lambda, ~~~ V_{cb}/V_{us} \sim \epsilon, ~~~
V_{ub}/V_{cb} \sim \lambda. \ee

In Eqs.~(\ref{quarktrial-01}) and (\ref{quarktrial-02}) of the
Appendix we present two examples of corrections which reproduce
all parameters of the quark sector. Notice that indeed, the
inequalities $\epsilon_{ij}^u,$ $\epsilon_{ij}^d < \lambda$ are
satisfied for all $i,j$. Both up and down matrices contain some
elements of the order $\lambda$. Some corrections are much smaller
than $\lambda$. Furthermore, two examples have different dominant
structures (sets of matrix elements of the order $\lambda$). The
detailed study of properties of $\epsilon_{ij}$ will be given
elsewhere~\cite{DS}.

\subsection{Lepton mixing: flipping the sign of rotation}
\label{flip}

In our approach an enhancement of the lepton mixing is a
consequence of the seesaw mechanism. The seesaw produces two
effects:
\begin{enumerate}
    \item It flips the sign of rotation which diagonalizes the mass matrix
of light neutrinos $\hat{m}_{\nu}$ with respect to the sign of the
rotations which diagonalize the Dirac neutrino matrix
$\hat{m}_{D}$ and charged lepton mass matrix. As a result, the
rotations of the neutrinos and charged leptons sum up in the
lepton mixing matrix;
    \item It enhances moderately the mixing produced by the
neutrino
mass matrix.
\end{enumerate}

Let us consider these effects for the 2-3 mixing explicitly.
Diagonalizing the 2-3 submatrix of $\hat{m}_{\nu}$ we find \be
\tan 2\theta^{\nu}_{23} = 2 \lambda \left[(1 +  \epsilon_{23}^{D})
+ \frac{(\epsilon_{23}^{D} - \epsilon_{23}^{M})(\epsilon_{22}^{D}
- 2 \epsilon_{23}^{D}-\epsilon_{23}^{D~2})}{\epsilon_{22}^{M} - 2
\epsilon_{23}^{M} -2 \epsilon_{23}^{D} \epsilon_{23}^{M}+
\epsilon_{23}^{D~2} + \lambda^2 \cdot O(\epsilon)} \right].
\label{theta} \ee The first term in square brackets corresponds to
diagonalization of the Dirac mass matrix; the second one is the
effect of seesaw. An explanation of the magnitude of the 2-3
mixing requires the second term to be $\sim -3$. So that in
combination with the first term it gives $\tan 2\theta^{\nu}_{23}
\sim - 4 \lambda$.

Notice that the seesaw contribution is proportional to the
difference of the off-diagonal (2-3) corrections and,
approximately, the ratio of determinants of the Dirac and Majorana
neutrino mass matrices. Since the determinants equal the
corresponding mass hierarchies, the enhancement of mixing requires
much stronger hierarchy of the RN neutrino masses than hierarchy
of the eigenvalues of the Dirac matrix.

This can be seen explicitly by considering the mass matrix of
light neutrinos:
\begin{equation}
\hat{m}_{\nu} \sim \frac{1}{\epsilon_{22}^M - 2\epsilon_{23}^{M} -
\epsilon_{23}^{M~2}} \left(
\begin{tabular}{ll}
$A_{22}\lambda^2$ & $A_{23}\lambda$\\
$A_{23}\lambda$ & $A_{33}$\\
\end{tabular}
\right),
\label{mat}
\end{equation}
where $A_{ij} \equiv A_{ij}(\epsilon_{kl}^D, \epsilon_{kl}^M)$. We
find explicitly that \be A_{ij} = \epsilon_{22}^M - 2
\epsilon_{23}^M + O(\epsilon_{ij}^2). \label{defA} \ee That is,
the coefficients $A_{ij}$ are all equal to each other in the
lowest (first) order in $\epsilon_{ij}$. Therefore to enhance the
mixing and to flip the  sign of rotation the terms of the order
$\epsilon^2$ in (\ref{defA}) should be important. Consequently,
\be \epsilon_{22}^M = 2 \epsilon_{23}^M + O(\epsilon_{ij}^2)
\label{cancel} \ee and $A_{ij} = O(\epsilon_{ij}^2).$ The equality
(\ref{cancel}) means that the determinant of the Majorana matrix
of the RH neutrino components is of the order $\lambda^2 \cdot
O(\epsilon_{ij}^2)$ or smaller, and consequently, the RH neutrino
masses have strong hierarchy: \be \frac{M_2}{M_3} \sim \lambda^2
\cdot \epsilon^2 \leq \lambda^4, \ee whereas $m_2^D/m_3^D \sim
\lambda^2\epsilon$. It is this difference of hierarchies which
leads to the seesaw enhancement of the 2-3 mixing.

There are three different possibilities to realize the flip of the
sign of the neutrino rotation:
\begin{enumerate}
    \item Change the sign of the off-diagonal mass terms $(\hat{m}_\nu)_{23}$.
    \item Change the sign of the diagonal mass term  $(\hat{m}_\nu)_{33}$
(provided that  $|(\hat{m}_\nu)_{33}| > |(\hat{m}_\nu)_{22}|$).
    \item Enhance the 22 element, so that  $(\hat{m}_\nu)_{22} > (\hat{m}_\nu)_{33}$.
\end{enumerate}
In terms of Eq.~(\ref{theta}) the sign of the second (seesaw) term
can be changed in four different ways by appropriately changing
the sign of the factors in its numerator and/or denominator.
Numerically, we find this to happen in 5 \% of cases for randomly
generated coefficients $\epsilon^M_{22}$, $\epsilon^M_{23}$,
$\epsilon^D_{22}$ and $\epsilon^D_{23}$ in the allowed range given
in (\ref{eps}).

Summarizing, generically, the mass matrix of the left-handed (LH)
neutrinos has the form (\ref{mat}) with moderately enhanced
off-diagonal term: $|A_{23}/A_{33}| \sim 2~$--$~3$. The relative
sign of $A_{23}$ and $A_{33}$ is negative. In large region of
parameter space $A_{22}$ can be comparable with two other
elements. That corresponds to summing up different (order
$\epsilon^2$) contributions, thus producing not too strong mass
hierarchy.

\subsection{Seesaw and the 1-2 neutrino sector}
\label{IIIE}

The mass matrix of the light neutrinos can be written as
\begin{equation}
\hat{m}_{\nu} = - U_L \hat{m}_D^{diag} V
(\hat{M}_R^{diag})^{-1}V^T \hat{m}_D^{diag} U^T_L~~, \label{mnure}
\end{equation}
where \be V = U_{R}^{\dagger} U_M =
U_{R12}^{\dagger}U_{R13}^{\dagger}U_{R23}^{\dagger} U_{M23}
U_{M13} U_{M12}, \label{vmatr} \ee and $U_{R}$ and $U_M$ are the
rotations of the RH neutrino components which diagonalize
$\hat{m}_D$ and $\hat{M}_R$ correspondingly.

We find in the lowest order in $\lambda$ and $\epsilon$:
\begin{equation}
V \approx \left(
\begin{tabular}{ccc}
$\cos \Delta_{12}$    &  $~\sin \Delta_{12}~$ &
$~\cos \theta_{12}^R\sin \Delta_{13}- \sin \theta_{12}^R\sin \Delta_{23}$ \\
$- \sin\Delta_{12}$  &  $\cos\Delta_{12}$ & $\sin \Delta_{23}$ \\
$-\cos\theta_{12}^M \sin\Delta_{13} + \sin\theta_{12}^M
\sin\Delta_{23}~$ &
$~- \sin \Delta_{23}~$  & 1 \\
\end{tabular}
\right),
\label{Vmatr}
\end{equation}
where $\Delta_{ij} \equiv \theta_{ij}^M - \theta_{ij}^R$, and the
angles $\theta_{ij}$ are determined in Eqs.~(\ref{angle23}),
(\ref{angle13}) and (\ref{angle12}).

Due to equality $\epsilon^M_{22} \approx 2\epsilon^M_{23}$,
according to (\ref{angle12}) the angle $\theta_{12}^M$ can be near
$\pi/4$, so that $\sin \Delta_{12} \sim 1$, $\cos \Delta_{12} \sim
\lambda~$--$~1$, $\sin \Delta_{23} = O(\lambda\epsilon)$, $\sin
\Delta_{13} = O(\lambda^2\epsilon)$ and $\sin \theta_{12}^R =
O(\lambda)$. Using these estimations we find
\begin{equation}
V \approx \left(
\begin{tabular}{ccc}
$ \lambda$ & 1 &  $ \Delta_{13} -\Delta_{23} \lambda$  \\
$ - 1 $        &  $ \lambda$     & $~\Delta_{23}$ \\
$\Delta_{23}~$  &  $~-\Delta_{23}~$ & 1 \\
\end{tabular}
\right), \label{Vappr}
\end{equation}
where $\Delta_{23}=(\epsilon^M_{23}-\epsilon^D_{23}) \lambda$ and
$\Delta_{13}=(\epsilon^M_{13}-\epsilon^D_{13}) \lambda^2$. Taking
the hierarchy of the mass eigenvalues as $\hat{M}^{diag}_{R} \sim
(\epsilon \cdot \lambda^4 ,\epsilon^2 \cdot \lambda^2 , 1)$ and
$\hat{m}^{diag}_{D} \sim (\lambda^4 , \epsilon \cdot \lambda^2,
1)$ we find from (\ref{Vappr}) an estimate of the light neutrinos
mass matrix (before the LH rotations):
\begin{equation}
\hat{m}_D^{diag} V (\hat{M}_R^{diag})^{-1}V^T \hat{m}_D^{diag}
\approx \left(
\begin{tabular}{ccc}
$\lambda^6/\epsilon^2$ &  $-\lambda^3$
& $~\Delta_{23} \frac{\lambda}{\epsilon} (1-\frac{\lambda}{\epsilon})$  \\
$-\lambda^3$ &  $\epsilon$
& $-\Delta_{23}/\lambda^2$ \\
$\Delta_{23} \frac{\lambda}{\epsilon}
(1-\frac{\lambda}{\epsilon})~$ & $~-\Delta_{23}/\lambda^2~$
& 1 \\
\end{tabular}
\right) m_3. \label{mappr}
\end{equation}
Note that $\Delta_{13}$ does not contribute in the leading order
in $\lambda$. If we set $\Delta_{23}\approx\lambda^2$ the light
neutrino mass matrix in the flavor basis with the LH rotating
included takes the form
\begin{equation}
\hat{m}_\nu \approx \left(
\begin{tabular}{ccc}
$\epsilon \lambda^2~$ &  $~\epsilon \lambda~$
& $~\lambda$  \\
$\epsilon \lambda$ &  $\epsilon \lambda $
& $1$ \\
$\lambda$ & $1$
& 1 \\
\end{tabular}
\right) m_3, \label{mappr1}
\end{equation}
where we show only the leading terms in both $\epsilon$ and
$\lambda$. Notice that in the 12 element of (\ref{mappr1}) the
combination $\epsilon$ should be enhanced to generate large 1-2
mixing. This will lead, after the 2-3 rotation, to the 1-3 term of
the order $\lambda^2$ according to our general considerations.

\section{Phenomenological consequences}
\label{predictions}

The corrections $\epsilon_{ij}^K$ have been introduced in a
certain way (\ref{ytotal}) and they are restricted to be small
enough (\ref{eps}). This allows us to draw some qualitative
consequences. Though exact predictions would require determination
of $\epsilon_{ij}^K$.

For illustration, in the Table~\ref{tab:table2} we present two
examples of the matrices of corrections. They correspond to two
different realizations of the sign flip: the Example I(l)
implements the inequality $(\hat{m}_\nu)_{22} >
(\hat{m}_\nu)_{33}$ and in the Example II(l) the element
$(\hat{m}_\nu)_{23}$ changes the sign. For simplicity we take
$\epsilon_{13} =  \epsilon_{11} = 0$. With these corrections the
mass matrices reproduce precisely the lepton mixings, charged
lepton masses  and the neutrino mass squared differences.

The predictions from these two sets of matrices are given in the
Table~\ref{tab:table3} where we present values of the lightest
neutrino mass, the effective Majorana mass of the electron
neutrino, the value of $U_{e3}$ and masses of the RH neutrinos.
Since the neutrino mass spectrum is hierarchical the radiative
corrections are very small \cite{Babu:qv,Frigerio:2002in}.

\subsection{1-3 mixing}

Generically for the 1-3 mixing we expect $U_{e3} \sim \lambda^2
\approx 0.07$. If $\epsilon^K_{13} = 0$ ($K=l,D,M$), we find \be
U_{e3} = \sin \theta_{13}^{\nu} - \sin \theta_{13}^{l} \cos
\theta_{23} - \sqrt{\frac{m_e}{m_{\mu}}} \sin \theta_{23} +
O(\lambda^4), \label{mix13} \ee where $\theta_{13}^{\nu}$ and
$\theta_{13}^{l}$ are the angles of rotations which diagonalize
the neutrino and charged lepton mass matrices, and $\theta_{23}
\equiv \theta^\nu_{23}-\theta^l_{23}$. In (\ref{mix13}) the last
term is induced by simultaneous 12 and 23 rotations. Notice that
in the sum each contribution is of the order $\lambda^2$ and the
next order correction is very small. So, depending on sign of the
angle and phase one may get substantial cancellation of the terms,
and even $U_{e3} = 0$ can be achieved. If however the terms sum up
we can get $U_{e3} \sim 0.2$ which corresponds to the present
upper experimental bound.

We also refer to the results of the numerical analysis summarized
in the Table~\ref{tab:table3}. (For details on numerical procedure
see the Appendix.) In the examples considered in the Appendix,
$|U_{e3}|^2$ is indeed of the order $\lambda^4$.
\begin{table}[h]
\caption{\label{tab:table3} Majorana masses of the RH neutrinos,
$m_{ee}$, $m_1$ and predicted value of $|U_{e3}|$. To extract the
value of the RH neutrino masses we take $m_3 =0.045$\,eV. $M_{i}$
are given in GeV's, whereas the masses $m_{ee}$ and $m_1$ are in
eV's.}
\begin{ruledtabular}
\begin{tabular}{lcccccc}
& $M_1$ &$M_2$ &$M_3$ &
$m_{ee}$ & $m_1$ & $|U_{e3}|^2$\\
\hline Example I(l) & $1.3 \times 10^{10}$ & $3.0 \times 10^{10}$&
$8.6 \times 10^{14}$&
$0.0006$&  $0.002$& $0.008$\\
Example II(l) &  $2.5 \times 10^{8}$ & $2.2 \times 10^{11}$& $3.8
\times 10^{14}$&
$0.0007$& $0.004$& $0.001$\\
\end{tabular}
\end{ruledtabular}
\end{table}

\subsection{The absolute scale of neutrino mass and $m_{ee}$}

According to our general consideration in Section~\ref{IIIE}, the
spectrum of light neutrinos is hierarchical, so that numerically
$m_3$ and $m_2$ are determined by the mass squared differences
measured in the atmospheric and solar neutrino experiments: $m_3
\approx \sqrt{\Delta m^2_{atm}} \cong 0.045$\,{eV} and $m_2
\approx \sqrt{\Delta m^2_{sol}}$. Parametrically $m_2/m_3 =
\lambda^2 \epsilon$ which implies that $\epsilon =  2.7$. The
lightest mass can be found evaluating the determinants of the
matrices in (\ref{eq:seesaw}). Indeed, parametrically
$\mathrm{Det}\,(\hat{Y_K})=\lambda^6\cdot\epsilon_K^2$, where
$\epsilon_K$ represents linear combinations of $\epsilon^K_{ij}$
coefficients, so that the determinant of the seesaw matrix
$\mathrm{Det}\,(\hat{m}_\nu) = \lambda^6
\epsilon_D^4/\epsilon_M^2$. Then, we have: \be m_1 =
\frac{\mathrm{Det}\,(\hat{m}_{\nu})}{m_2 m_3} = \lambda^4
\frac{\epsilon_D^4}{\epsilon \epsilon_M^2} m_3,\ee where
$\epsilon_D=\epsilon_D(\epsilon^D_{ij})$,
$\epsilon_M=\epsilon_M(\epsilon^M_{ij})$ and $\epsilon=2.7$. In
the examples presented in the Table~\ref{tab:table2} the hierarchy
of light masses is rather weak: $m_1/ m_2 = 0.2~$--$~0.5$ which is
partly related to strong hierarchy of the RH neutrino masses. For
the lightest mass we get (see the Table~\ref{tab:table3})
typically \be m_1 \sim (0.1~\mbox{--}~5) \times 10^{-3}\,{\rm eV}.
\ee Taking this into account we obtain $\epsilon_D^4/\epsilon_M^2
\sim (1~$--$~10^2)$ which can be used to estimate how singular
$\hat{Y}_D$ and $\hat{Y}_M$ are with respect to each other.

The effective Majorana mass of the electron neutrino can be
calculated immediately as the $ee$-element of the neutrino mass
matrix in the flavor basis (\ref{mappr1}). Parametrically this
gives $m_{ee} = \epsilon \lambda^2 m_3$, where $\epsilon$ stands
for the linear combination of $\epsilon_{ij}$'s.

Alternatively, we can use the neutrino masses and known neutrino
mixing and present $m_{ee}$ as the sum of contributions of mass
eigenstates:
\begin{equation}\label{}
    m_{ee}=\sum_i |U_{ei}|^2 m_i e^{i \phi_i}=m_{ee}(1)+m_{ee}(2)+m_{ee}(3).
\end{equation}

In general, due to smallness of the 1-3 mixing the contribution of
$\nu_3$ is very small: $m_{ee} (3) = \sqrt{\Delta m^2_{atm}}
\lambda^4 \sim 2 \times 10^{-4} $\,eV. The contribution of $\nu_2$
is phenomenologically determined: $m_{ee} (2) = \sqrt{\Delta
m^2_{sol}} \sin^2\theta_{sol} \sim (2~\mbox{--}~3) \times
10^{-3}$\,eV, and usually dominates. The contribution of $\nu_1$
can be comparable with the previous one due to weak mass hierarchy
and larger admixtures of $\nu_e$. Furthermore, typically the
masses and therefore contributions of $\nu_1$ and $\nu_2$ have an
opposite sign cancelling each other in $m_{ee}$. For this reason
the predictions for $m_{ee}$ in the examples of the
Table~\ref{tab:table2} are small: $m_{ee} \sim 10^{-3}$\,eV.

If the Heidelberg-Moscow positive
result~\cite{HM-neg1,HM-pos,vissani,HM-neg2,Klapdor-Kleingrothaus:2004ge}
is confirmed, either our approach, at least in its present form,
is not correct or another mechanism, different from the Majorana
mass of the light neutrinos gives main contribution to the decay
rate.

\subsection{Leptogenesis}

The corrections $\epsilon^K_{ij}$ are in general complex numbers
and this is the source of CP violation in our approach.

Since $M_{3} \gg M_{2}, M_1$, only two lighter RH neutrinos are
relevant for leptogenesis and the lepton number asymmetry can be
written
as~\cite{Luty:un,Flanz:1994yx,Plumacher:1996kc,Covi:1996wh,Buchmuller:1997yu,Giudice:2003jh}
\be \epsilon_{L} =  \frac{1}{8\pi} \frac{M_1}{M_2}
\frac{(h^{\dagger} h)_{12}^2}{(h^{\dagger} h)_{11}}. \ee Here $h$
is the matrix of the Yukawa couplings in the basis where
$\hat{M}_R$ is diagonal. Apparently, \be h^{\dagger} h = V^T
\frac{\hat{m}_D^{diag~2}}{v_1^2} V, \ee where $V$ is determined in
(\ref{Vappr}). Using estimations of the matrix elements of $V$
(and assuming that the imaginary parts of these elements can be as
large as the real ones) we find the asymmetry \be \epsilon_{L} =
\frac{\lambda^5}{8\pi} \sim 5 \times 10^{-5}. \ee Then the baryon
to photon ratio is given by \be \eta_B \sim 0.01 \epsilon_{L} k_1,
\ee where $k$ describes the washout of the produced lepton
asymmetry due to weak deviation from the thermal equilibrium. The
factor $k$ depends on the effective mass parameter \be \tilde{m}_1
= \frac{v_1^2(h^{\dagger} h)_{11} }{M_{1}} \sim
\mbox{(0.1~--~1)\,eV} . \label{mtilde} \ee For this value of the
effective mass we get $k_1 \sim 10^{-3}$, and therefore $\eta_B
\sim 5 \times 10^{-10}$ in agreement with the observed value.

Notice that the key difference of our scenario from the analysis
in \cite{rhma} is that the lightest eigenvalue of the neutrino
Dirac matrix is much larger than the up quark mass $m_u$; in the
Example I(l): $m_{1D} \sim 300$\,MeV. Also, the left rotations are
not negligible here.

\subsection{Lepton number violating effects}
\label{LNVE}

In the SUSY context one expects observable flavor violating
decays, like $\mu \rightarrow e \gamma$, due to slepton mixing
related to the neutrino mixing. The approximate formula for the
$\mu \rightarrow e \gamma$ branching ratio, which has the most
stringent experimental limit, reads
\begin{equation}
BR(\mu \rightarrow e \gamma)\simeq \frac{\alpha^3}{G_F^2}
\frac{|(\delta m^2_{\tilde{L}})_{21}|^2}{m_s^8} \tan^2 \beta,
\end{equation}
where $m_s$ stands for the effective  mass of the superparticles,
and $\delta m^2_{\tilde{L}}$ represents the off-diagonal
corrections to the slepton mass matrix. They appear due to the
renormalization group running between the scale where universality
conditions on SUSY breaking parameters are imposed, which we take
to be the GUT scale $M_{GUT}$, and scale where the RH neutrinos
decouple from the theory.

We find
\begin{equation}
(\delta m^2_{\tilde{L}})_{ij} \approx \frac{3 m_0^2+A_0^2}{8
\pi^2} (V_{L})_{i3} (\hat{Y}_D^{diag})_{33} (V^\dag_{L})_{3j} \ln
\frac{M_{GUT}}{M_3},
\end{equation}
where the matrix $V_{L}=U_l^\dag U_L$ represents the mismatch in
the LH rotations that diagonalize $\hat{Y_l}$ and $\hat{Y_D}$. The
relevant coefficients $(V_{L})_{13}$ and $(V_{L})_{23}$ for the
$\mu \rightarrow e \gamma$ process are proportional to $\lambda^2
\cdot O(\epsilon)(\leq \lambda^3)$ and $\lambda \cdot
O(\epsilon)(\leq \lambda^2)$ respectively. Though, the exact
values depend on combinations of $\epsilon$, we expect the product
$(V_{L})_{13} (V_{L})_{23}$ to be close to $V_{ub}
V_{cb}\leq\lambda^5$. (See also the form of $V$ in
Eq.~(\ref{Vmatr}) and the discussion on the mixing in the quark
sector.)

Rather precise approximation for the effective mass $m_s$ is given
by \cite{Petcov:2003zb}
\begin{equation}
m^8_s\simeq 0.5 m^2_0 M^2_{1/2} (m^2_0+0.6 M^2_{1/2})^2,
\end{equation}
where $m_0$ is the typical slepton mass and $M_{1/2}$ is the
gaugino mass. Taking for simplicity, $m_0=M_{1/2} \equiv m$ and
value $(V_{L})_{13} (V_{L})_{23}=\lambda^5$ we obtain
\begin{equation}
BR(\mu \rightarrow e \gamma) \simeq 1.8 \times 10^{-9}
\left(\frac{m}{100\,{\rm GeV}}\right)^{-4}, \label{br1}
\end{equation}
where $\lambda=0.26$, $\tan \beta =55.9$,
$(\hat{Y}_D^{diag})_{33}\simeq0.7$, $M_{GUT}/M_3=100$ and $A_0 =
0$ were used. In the case of an exact quark-lepton symmetry:
$(V_{L})_{13} (V_{L})_{23}=V_{ub} V_{cb}$ we find
\begin{equation}
BR(\mu \rightarrow e \gamma) \simeq 1.1 \times 10^{-11}
\left(\frac{m}{100\,{\rm GeV}}\right)^{-4}. \label{br2}
\end{equation}
According to Eqs.~(\ref{br1}) and (\ref{br2}) for $m \simeq
(300~$--$~400)$\,GeV we expect $BR(\mu \rightarrow e \gamma)
\simeq 10^{-13}~$--$~10^{-11}$. This interval is close to the
current experimental limit of $BR(\mu \rightarrow e \gamma) < 1.2
\times 10^{-11}$ \cite{Hagiwara:fs} and clearly within reach of
the MEG experiment at PSI \cite{Mori:sg} which will have a
sensitivity down to $BR(\mu \rightarrow e \gamma) \leq 5 \times
10^{-14}$.

The results of an exact numerical running\footnote{Due to the
highly non-degenerate spectrum of the RH neutrinos the care has
been taken to integrate them out at the appropriate energy scales
as suggested in \cite{Antusch:2002rr}.} of slepton mass matrix are
also in an excellent agreement with the approximations presented
in this section.

Finally, we note that the majority of the SUSY GUT models yields
significantly larger value for the product $(V_{L})_{13}
(V_{L})_{23}$ than what is generated in our approach. This puts
them in precarious position with respect to the experimental
constrains on lepton flavor violating processes. Namely, they
typically yield $(V_{L})_{13} (V_{L})_{23} \sim
10^{-2}$--$10^{-1}$ (see for example \cite{Barr:2003fn} and
references therein) which makes them violate the experimental
bounds even for low values of $\tan \beta$ ($\sim 5$). On the
other hand generically we obtain $(V_{L})_{13} (V_{L})_{23} \sim
10^{-4}~$--$~10^{-3}$, which can be traced back to the ansatz
(\ref{yzero}). This rather large suppression more than compensates
the enhancement of $\mu \rightarrow e \gamma$ branching ratio that
originates from the large value of $\tan \beta$. The suppression
brings our prediction for $\mu \rightarrow e \gamma$ branching
ratio close to but below the current experimental limit.

\section{Discussion and Implications}
\label{discussion}

\noindent 1). There are two different approaches to the theory of
fermion masses. One possibility (widely explored in the
literature) is to build up the theory immediately on the basis of
observables---the mixing and mass ratios---considering them as
fundamental parameters. In this case the quark-lepton symmetry is
{\it strongly\/} broken at least by masses of the first and second
generations. In a number of models this is described by
introduction of different charges for the leptons and quarks.
Another approach is when the quark-lepton symmetry is {\it
weakly\/} broken. In this case, the observables appear as
diagonalization of the nearly equal mass matrices. They are
determined by small corrections to the dominant structure equal
for quarks and
leptons.\\

\noindent 2). The main feature of our approach is the nearly
singular matrices $\hat{Y}_K$. This allows us, using small
perturbations, to generate strong difference of the mass
hierarchies of quarks and leptons and simultaneously enhance the
lepton mixing. The lepton mixing (due to the seesaw mechanism) is
unstable with respect to perturbation of the RH mass matrix which
appears in the denominator of the expression for the light
neutrino masses. The perturbations of $M_R$ influence strongly the
mass hierarchy of the RH neutrinos and therefore (via the seesaw
enhancement) the
mixing of light neutrinos.\\

\noindent 3). The matrix $\hat{Y}_0$ (\ref{yzero}) can be obtained
in the model with U(1) family symmetry in the context of the
Froggatt-Nielsen~\cite{FN} (F-N) mechanism. According to this
mechanism the Yukawa couplings are generated by the operators
\begin{equation}\label{addition}
    a_{ij} f^{c}_{iL} f_{jL} \left( \frac{\sigma}{M_F}\right)^{q_i+q_j}
    H_k,
\end{equation}
where $f_i$ are fermion components, $a_{ij}$ are dimensionless
constants of order 1, and $\sigma$ is the scalar field---singlet
of the SM gauge symmetry group with a $U(1)_F$ charge of $-1$.
$H_k$ $(k=1,2)$ are the Higgs doublets of the MSSM, $M_F$
corresponds to mass scale at which the non-renormalizable
operators describing interactions of $\sigma$ with fermion fields
are generated and $q_i$ ($i = 1, 2, 3$) are the $U(1)_F$ charges
of the family $i$.

After $\sigma$ develops the VEV $\langle \sigma \rangle$ the
following Yukawa couplings are generated: \be (\hat{Y})_{ij} =
a_{ij} \lambda^{q_i + q_j},\quad \lambda = \frac{\langle \sigma
\rangle}{M_F}. \label{FNcharge} \ee If $a_{ij}=a_0=1$, the matrix
$\hat{Y}=\hat{Y}_0$ is singular. Furthermore, prescribing the
$U(1)_F$ charges $0$, $1$, and $2$ for the third, second, and
first family we reproduce the required structure of matrix
$\hat{Y}_0$ (\ref{yzero}). Corrections to $Y_0$ can be generated
by deviations of $a_{ij}$ from universality:
\begin{equation}\label{a}
    a_{ij}=a_0 (1+\epsilon_{ij}).
\end{equation}

In general, the required singular matrix can be represented as the
product: \be \hat{Y}_0 =  W \times W^T, ~~~~ W^T \equiv (a_1, a_2,
a_3). \label{dom} \ee In turn, such a structure can appear as a
result of interaction of the light fermion fields $(f_1, f_2,
f_3)$ with a single heavy field $F$. Let us consider the following
mass terms: \be \bar{F}_L \sum_{i = 1}^{3} \mu_i f_{iR} +
\bar{F}_R \sum_{i = 1}^{3} \mu_i f_{iL} + h.c. \label{massterm}
\ee with $\mu_i < M$ and the Dirac mass terms formed by $f_{iL}$
and $f_{iR}$ are forbidden by some symmetry. Then after decoupling
of $F$ we get for the light masses \be m_{ij} = \frac{\mu_i
\mu_j}{M} \ee with required properties. Notice however, that this
mechanism cannot be applied immediately to top
quark since $m_t \sim \mu_i|_{max}.$\\

\noindent 4). To reproduce observables we still need small
deviations of coefficients $a_{ij}$ from 1. This may come from the
F-N mechanism itself as it is indicated in (\ref{a}) or from new
physics at some higher scale as an additional contribution to mass
matrices. The correction matrix is of the order
$\lambda^2~$--$~\lambda^3 \sim (1~$--$~3) \times 10^{-2}$. So, if
the flavor symmetry is realized at the GUT scale the correction
matrix can be related to some physics at the string
scale.\\

\noindent 5). The case of unstable matrices reproduces to some
extent a situation of anarchy: small perturbations of the
otherwise symmetric pattern lead to significant difference in the
observables.

Selecting $a_i$ in (\ref{dom}) one can further ``optimize'' the
structure of the dominant singular matrix to reduce spread of the
the corrections $\epsilon_{ij}^K$, to diminish their absolute
values or to get certain relations~\cite{DS}.\\

\noindent 6). To explain the observed masses and mixing certain
relations between the correction parameters $\epsilon$ should be
satisfied and some of them should be in narrow ranges. These
relations should be used to construct the theory of $\epsilon$.
Random selection of values of $|\epsilon|$'s in the intervals
$0~$--$~\lambda$ produces typically incorrect values of the
observables.

\section{Conclusions}
\label{conclusions}

We have elaborated an approach in which no {\it ad hoc} symmetry
for the neutrino sector is introduced. The difference of
parameters in the quark and the lepton sectors arises essentially
from the seesaw mechanism as well as from ``instability'' of the
mass matrices. The difference of the mass hierarchies follows from
small perturbations of the singular matrices. Singularity can be a
consequence of certain family symmetry.

The explanation of features of observables is reduced to large
extent to explanation of perturbations (corrections). Particular
values of $\epsilon$'s are needed. Still our proposal opens
alternative approach to explain the data. Furthermore, in this
approach one gets
\begin{enumerate}
\item correct hierarchy of the quark mixings;
    \item hierarchical mass spectrum of light neutrinos;
    \item 1-3 mixing of the order $\lambda^2$;
    \item small effective Majorana mass of the electron neutrino:
    $m_{ee}\leq 10^{-2}$\,eV;
    \item in general, deviation of the 2-3 mixing from maximal;
    \item generic prediction is a strong mass hierarchy of the second
and third  RH neutrinos which is of the order $\lambda^4$.
\end{enumerate}

Perturbations of the singular matrix introduced in the form
(\ref{ytotal}) with $|\epsilon_{ij}| \leq \lambda = 0.26$ allow us
to reproduce all available experimental results. Even in
parametric form the approach leads to correct qualitative pattern
of masses and mixings though quantitative description of the data
requires precise determination of $|\epsilon_{ij}|$ within the
interval $0~$--$~\lambda$. Let us summarize the information on
$\epsilon$ we have obtained:

\begin{itemize}
    \item We have shown that the data can be well described for all
$\epsilon \leq 0.26$.
    \item There is rather strict relation (\ref{cancel}) required by the
enhancement of the 2-3 leptonic mixing. The same relation also
gives enhancement of the 1-2 mixing.
    \item $\epsilon_{23}$ for the Dirac mass matrices are determined by
the mass ratios for the second and third generations.
    \item $\epsilon_{22}$ elements correlate with $\epsilon_{23}$ and
they are restricted by the 2-3 CKM mixing in the quark sector.
    \item Values of $\epsilon_{11}$ are practically irrelevant.
    \item There are rather complicated relations between other
parameters (they also include parameters of the 2-3 sector) which
follow from masses of first generation, the 1-2 leptonic mixing
and CKM mixing. These relations do not restrict a given parameter
once other parameters are allowed to change in the intervals $|0 -
\lambda|$.
    \item The description of all available data leaves substantial
freedom of variations of these parameters ($\epsilon_{12}$,
$\epsilon_{13}$). So one can impose on them additional conditions
motivated by theoretical context (zeros, equalities, {\it etc}.)
\cite{DS}.
\end{itemize}

\section{Acknowledgments}

We would like to thank Z.~Berezhiani, R.~Dermisek, W.~Liao and
Y.~Takanishi for fruitful discussion.

\appendix*
\section{}
We present the numerical results for $\epsilon^K_{ij}$
corrections. Input parameters, the masses and mixings of the
matter fields, at the GUT scale used for the numerical fit, are
given in the Table~\ref{tab:table1}. We assume the MSSM particle
content below the GUT scale and determine $\tan \beta$ requiring
the unification of $b$ and $t$ Yukawa couplings.
\begin{table}[h]
\caption{\label{tab:table1} Experimental values of the quark and
charged lepton masses and relevant CKM angles extrapolated to the
GUT scale. Three-loop QCD and one-loop QED renormalization group
equations are used in running up to $m_t$. Further extrapolation
from $m_t$ to $M_{GUT}=2.3 \times 10^{16}$\,GeV is done using the
two-loop MSSM beta functions taking all SUSY particles to be
degenerate at $m_t$ and assuming $\tan \beta = 55.9$. Masses are
given in GeV.}
\begin{ruledtabular}
\begin{tabular}{cccccccccccc}
$m_u$&$m_c$&$m_t$&$m_d$&$m_s$&$m_b$&$m_e$&$m_{\mu}$&$m_{\tau}$&
$|V_{us}|$&$|V_{ub}|$&$|V_{cb}|$\\
\hline $0.000558$ & $0.264$ & $121$ & $0.00137$ & $0.0239$ &
$2.16$ & $0.000530$ & $0.110$ & $2.45$ & $0.222$ & $0.00284$ &
$0.0320$
\end{tabular}
\end{ruledtabular}
\end{table}

In Eqs.~(\ref{quarktrial-01}) and (\ref{quarktrial-02}) we present
two examples of correction matrices of quarks which yield exact
agreement with the experimental input in the
Table~\ref{tab:table1}.

\noindent Example I(q):
\begin{equation}\label{quarktrial-01}
\epsilon^u\cong\left(%
\begin{array}{ccc}
  0 & 0.0683 & -0.0103 \\
  0.0683 & 0.144 & 0.0526 \\
  -0.0103 & 0.0527 & 0 \\
\end{array}%
\right) \quad
\epsilon^d\cong\left(%
\begin{array}{ccc}
  0 & 0.0387 & -0.163 \\
  0.0387 & -0.00386 & -0.0821 \\
  -0.163 & -0.0821 & 0 \\
\end{array}%
\right)
\end{equation}

\noindent Example II(q):
\begin{equation}\label{quarktrial-02}
   \epsilon^u\cong \left(%
\begin{array}{ccc}
  0 & 0.00811 & -0.0100 \\
  0.00811 & 0.0200 & -0.00782 \\
 -0.0100 & -0.00782 & 0 \\
\end{array}%
\right) \quad
\epsilon^d\cong\left(%
\begin{array}{ccc}
  0 & -0.0112 & -0.160 \\
  -0.0112 & -0.110 & -0.141 \\
  -0.160 & -0.141 & 0 \\
\end{array}%
\right)
\end{equation}
The coefficients $(y_u, y_d)$ in the examples I(q) and II(q) are
$(0.645,0.655)$ and $(0.650,0.659)$ respectively. This difference
can also be accounted as $\lambda^2$ correction to $(33)$
elements. The parameter $\lambda$ is always 0.26.

We next specify two examples in the lepton sector in the
Table~\ref{tab:table2}. In the spirit of the simplest $SO(10)$
model we set $y_D=y_u\cong0.645$ for both cases. Our fit yields
$y_l=0.753$ in the Example I(l) and $y_l=0.754$ in the Example
II(l).
\begin{table}[h]
\caption{\label{tab:table2} Corrections $\epsilon^K_{ij}$ in the
lepton sector ($K=l,D,M$) for two different cases which realize
different scenarios for the flip of the sign of rotations. The fit
is performed assuming $m_2/m_3=0.187$ at the GUT scale. We also
require $\tan^2 \theta_{sol} = 0.4$ and $\sin^2 2 \theta_{atm} =
0.95$.}
\begin{ruledtabular}
\begin{tabular}{cc}
Example I(l) $((\hat{m}_\nu)_{33}<(\hat{m}_\nu)_{22})$ &
Example II(l) $((\hat{m}_\nu)_{23}<0,~(\hat{m}_\nu)_{33}>0)$\\
\hline &\\
$\epsilon^l\cong
\left(%
\begin{array}{ccc}
  0 & -0.171 & -0.036 \\
  -0.171 & 0.254 & -0.268 \\
  -0.036 & -0.268 & 0 \\
\end{array}%
\right)$ & $\epsilon^l\cong\left(%
\begin{array}{ccc}
  0 & -0.093 & 0.006 \\
  -0.093 & 0.262 & -0.262 \\
  0.006 & -0.262 & 0 \\
\end{array}%
\right)$ \\
&\\
$\epsilon^D\cong \left(%
\begin{array}{ccc}
  0 & 0.233 & 0 \\
  0.233 & 0.104 & 0.042 \\
  0 & 0.042 & 0 \\
\end{array}%
\right)$  & $\epsilon^D\cong\left(%
\begin{array}{ccc}
  0 & 0.213 & 0 \\
  0.213 & -0.200 & 0.098 \\
  0 & 0.098 & 0 \\
\end{array}%
\right)$ \\
&\\
$\epsilon^M\cong\left(%
\begin{array}{ccc}
  0 & 0.0065 & 0 \\
  0.0065 & 0.0098 & 0.005 \\
  0 & 0.005 & 0 \\
\end{array}%
\right)$
  & $\epsilon^M\cong\left(%
\begin{array}{ccc}
  0 & 0.130 & 0 \\
  0.130 & 0.264 & 0.129 \\
  0 & 0.129 & 0 \\
\end{array}%
\right)$ \\
&\\
\end{tabular}
\end{ruledtabular}
\end{table}

\newpage


\begin{thebibliography}{99}

\bibitem{Giunti:2003cq}
C.~Giunti,
arXiv:hep-ph/0309024.

\bibitem{Smirnov:2003xe}
A.~Y.~Smirnov,
arXiv:hep-ph/0311259.

\bibitem{HM-neg1} H.V.~Klapdor-Kleingrothaus {\it et al.},
{\it Eur. Phys. J.} A {\bf 12}, 147 (2001).

\bibitem{HM-pos} H.V.~Klapdor-Kleingrothaus  {\it et al.,}  {\it Mod. Phys. Lett.}
A {\bf 16},  2409 (2001).

\bibitem{vissani} F.~Feruglio, A.~Strumia, F.~Vissani,
{\it Nucl. Phys.} B {\bf 637},  345 (2002), {\it Addendum-ibid.},
B {\bf 659}, 359 (2003).

\bibitem{HM-neg2}
A.~M.~Bakalyarov, A.~Y.~Balysh, S.~T.~Belyaev, V.~I.~Lebedev and
S.~V.~Zhukov [C03-06-23.1 Collaboration],
arXiv:hep-ex/0309016.

\bibitem{Klapdor-Kleingrothaus:2004ge}
H.~V.~Klapdor-Kleingrothaus, A.~Dietz, I.~V.~Krivosheina and
O.~Chkvorets,
arXiv:hep-ph/0403018.




\bibitem{cosm3} S.~W.~Allen, R.~W.~Schmidt and S.~L.~Bridle, astro-ph/0306386.

\bibitem{cosm}D.~N.~Spergel {\it et al.},
{\it Astrophys. J. Suppl.}, {\bf 148}, 175 (2003),
[astro-ph/0302209].

\bibitem{cosm1} O.~Elgaroy, O.~Lahav, {\it JCAP}  {\bf 0304}, 004 (2003).

\bibitem{cosm2} S.~Hannestad, {\it JCAP}  {\bf 0305}, 004 (2003).



\bibitem{Low:2004wx}
C.~I.~Low,
arXiv:hep-ph/0404017.



\bibitem{A4}E.~Ma, G.~Rajasekaran, {\it  Phys.~Rev.} D {\it 64} 113012, (2001);
K.S.~Babu, E.~Ma, J.W.F.~Valle, {\it Phys. Lett.} B {\bf 552}, 207
(2003).

\bibitem{so3} R.~Barbieri, L.~J.~Hall, G.~L.~Kane and G.~G.~Ross, hep-ph/9901228.

\bibitem{Antusch:2004xd}
S.~Antusch and S.~F.~King,
arXiv:hep-ph/0402121.

\bibitem{Antusch:2004re}
S.~Antusch and S.~F.~King,
arXiv:hep-ph/0403053.

\bibitem{Altarelli:2003vk}
G.~Altarelli and F.~Feruglio,
arXiv:hep-ph/0306265.



\bibitem{atm} Super-Kamiokande Collaboration, Y.~Hayato, talk given at {\it the HEP2003
International Europhysics Conference} (Aachen, Germany, 2003),
website: eps2003.physik.rwth-aachen.de.

\bibitem{K2K} M.~H.~Ahn {\it et al.}, {\it Phys. Rev. Lett.} {\bf 90}:041801, (2003).

\bibitem{Weinberg:hb}
S.~Weinberg,
Trans.\ New York Acad.\ Sci.\  {\bf 38}, 185 (1977).

\bibitem{Wilczek:uh}
F.~Wilczek and A.~Zee,
Phys.\ Lett.\ B {\bf 70}, 418 (1977) [Erratum-ibid.\  {\bf 72B},
504 (1978)].

\bibitem{Fritzsch:1977za}
H.~Fritzsch,
Phys.\ Lett.\ B {\bf 70}, 436 (1977).

\bibitem{barshay}
S.~Barshay, G.~Kreyerhoff,  Lett.{\bf 63}, 519 (2003); S.~Barshay,
P.~Heiliger, Astropart. Phys. {\bf 6},  323 (1997).

\bibitem{sees} M.~Gell-Mann, P.~Ramond and R.~Slansky,
in {\it Supergravity}, eds P.~van Niewenhuizen and D.~Z.~Freedman
(North Holland, Amsterdam 1980); P.~Ramond, {\it  Sanibel talk},
retroprinted as hep-ph/9809459; T.~Yanagida, in {\it Proc.~of
Workshop on Unified Theory and Baryon number in the Universe},
eds. O.~Sawada and A.~Sugamoto, KEK, Tsukuba, (1979);
S.~L.~Glashow, in {\it Quarks and Leptons}, Carg\`ese lectures,
eds M.~L\'evy, (Plenum, 1980, New York) p. 707; R.~N.~Mohapatra
and G.~Senjanovi\'c, {\it Phys. Rev. Lett.} {\bf 44}, 912 (1980).

\bibitem{senhan} A.~Yu.~Smirnov, {\it Phys. Rev.} D {\bf 48} 3264 (1993);
M.~Tanimoto, {\it Phys. Lett.} B {\bf 345}, 477 (1995); T.K.~Kuo,
Guo-Hong Wu, Sadek W.~Mansour, {\it Phys. Rev.} D {\bf 61}, 111301
(2000); G.~Altarelli F.~Feruglio and I.~Masina, {\it Phys. Lett.}
B {\bf 472}, 382 (2000); S.~Lavignac, I.~Masina, C.~A.~Savoy, {\it
Nucl. Phys.} B {\bf 633}, 139 (2002); A.~Datta, F.~S.~Ling and
P.~Ramond, hep-ph/0306002; M.~Bando, {\it et al.,} hep-ph/0309310.

\bibitem{Akhmedov:2000yt}
E.~K.~Akhmedov, G.~C.~Branco, F.~R.~Joaquim and
J.~I.~Silva-Marcos,
Phys.\ Lett.\ B {\bf 498}, 237 (2001) [arXiv:hep-ph/0008010].

\bibitem{Dermisek:2003rw}
R.~Dermisek,
arXiv:hep-ph/0312206.

\bibitem{DS} I.~Dorsner and A.~Yu.~Smirnov, in preparation.


\bibitem{Babu:qv}
K.~S.~Babu, C.~N.~Leung and J.~Pantaleone,
Phys.\ Lett.\ B {\bf 319}, 191 (1993) [arXiv:hep-ph/9309223].

\bibitem{Frigerio:2002in}
M.~Frigerio and A.~Y.~Smirnov,
JHEP {\bf 0302}, 004 (2003) [arXiv:hep-ph/0212263].

\bibitem{Luty:un}
M.~A.~Luty,
Phys.\ Rev.\ D {\bf 45}, 455 (1992).

\bibitem{Flanz:1994yx}
M.~Flanz, E.~A.~Paschos and U.~Sarkar,
Phys.\ Lett.\ B {\bf 345}, 248 (1995) [Erratum-ibid.\ B {\bf 382},
447 (1996)] [arXiv:hep-ph/9411366].

\bibitem{Plumacher:1996kc}
M.~Plumacher,
Z.\ Phys.\ C {\bf 74}, 549 (1997) [arXiv:hep-ph/9604229].

\bibitem{Covi:1996wh}
L.~Covi, E.~Roulet and F.~Vissani,
Phys.\ Lett.\ B {\bf 384}, 169 (1996) [arXiv:hep-ph/9605319].

\bibitem{Buchmuller:1997yu}
W.~Buchmuller and M.~Plumacher,
Phys.\ Lett.\ B {\bf 431}, 354 (1998) [arXiv:hep-ph/9710460].

\bibitem{Giudice:2003jh}
G.~F.~Giudice, A.~Notari, M.~Raidal, A.~Riotto and A.~Strumia,
arXiv:hep-ph/0310123.



\bibitem{rhma}
E.~K.~Akhmedov, M.~Frigerio and A.~Y.~Smirnov,
JHEP {\bf 0309}, 021 (2003) [arXiv:hep-ph/0305322].




\bibitem{Petcov:2003zb}
S.~T.~Petcov, S.~Profumo, Y.~Takanishi and C.~E.~Yaguna,
Nucl.\ Phys.\ B {\bf 676}, 453 (2004) [arXiv:hep-ph/0306195].

\bibitem{Hagiwara:fs}
K.~Hagiwara {\it et al.}  [Particle Data Group Collaboration],
Phys.\ Rev.\ D {\bf 66}, 010001 (2002).

\bibitem{Mori:sg}
T.~Mori,
Nucl.\ Phys.\ Proc.\ Suppl.\  {\bf 111}, 194 (2002).

\bibitem{Antusch:2002rr}
S.~Antusch, J.~Kersten, M.~Lindner and M.~Ratz,
Phys.\ Lett.\ B {\bf 538}, 87 (2002) [arXiv:hep-ph/0203233].

\bibitem{Barr:2003fn}
S.~M.~Barr,
Phys.\ Lett.\ B {\bf 578}, 394 (2004) [arXiv:hep-ph/0307372].

\bibitem{FN}C.~D.~Froggatt and H.~B.~Nielsen, {\it Nucl. Phys.} B {\bf 147}, 277 (1979).

\end{thebibliography}
\end{document}